\documentclass[twocolumn,amsmath,amssymb,aps,prb,reprint,longbibliography]{revtex4-1}
\usepackage{graphicx}
\begin{document}

\title{
Drive Dependence of the Hall Angle for a Sliding Wigner Crystal in a Magnetic Field
} 
\author{
C. Reichhardt and C. J. O. Reichhardt 
} 
\affiliation{
Theoretical Division and Center for Nonlinear Studies,
Los Alamos National Laboratory, Los Alamos, New Mexico 87545, USA\\ 
} 

\date{\today}
\begin{abstract}
We numerically examine the depinning and sliding dynamics of a Wigner crystal in the presence of quenched disorder and a magnetic field. In the disorder-free limit, the Wigner crystal Hall angle is independent of crystal velocity, but when disorder is present, we find that Hall angle starts near zero at the depinning threshold and increases linearly with increasing drive before reaching a saturation close to the disorder free value at the highest drives. The drive dependence is the result of a side jump effect produced when the charges move over pinning sites. The magnitude of the side jump is reduced at the higher velocities. The drive dependent Hall angle is robust for a wide range of disorder parameters and should be a generic feature of classical charges driven in the presence of quenched disorder and a magnetic field.   
\end{abstract}

\maketitle
There are a wide range of systems containing quenched disorder
that exhibit depinning and sliding phenomena
\cite{Fisher98,Reichhardt17}, including vortices in type-II 
superconductors \cite{Bhattacharya93,Blatter94},
driven charge density waves \cite{Gruner88,Danneau02},
skyrmions \cite{Schulz12,Reichhardt15,Legrand17,Jiang17},
and colloids on substrates \cite{Reichhardt02,Pertsinidis08,Bohlein12}.
Such
systems can exhibit a threshold for motion and
nonlinear velocity-force curves as a function of increasing external
drive.
In many cases,
the depinning and sliding dynamics can be imaged directly
and compared to bulk transport measures,
showing that different types
of sliding phases are possible
including moving crystals and smectic states \cite{Reichhardt17}. 

A Wigner or electron crystal 
is also expected to exhibit  
depinning and sliding under an external drive
\cite{Andrei88,Goldman90,Williams91,Jiang91,Zhu94,Cha94,Chitra98,Perruchot98,Abrahams01,Reichhardt01,Reichhardt04,Zhang14,Jang17,Knighton18,Brussarski18}.
Experimentally,
the presence 
of nonlinear current-voltage curves or the onset of conduction noise
have been argued as
providing evidence for the depinning and
motion of Wigner crystals \cite{Goldman90,Williams91,Jiang91,Zhu94,Cha94,Perruchot98,Abrahams01,Reichhardt01,Reichhardt04,Brussarski18}; 
however, unlike superconducting vortices or
colloidal particles, direct visualization of the depinning
process is not possible, 
so additional transport measures for detecting the sliding  
of a Wigner crystal would be very valuable. 
If the sliding Wigner crystal is subjected to a magnetic field,
it moves at a Hall angle which is proportional to the strength of the
field, but relatively little is known about how the presence of 
a finite Hall 
angle could affect depinning or the sliding motion of a Wigner crystal.
Based on theoretical calculations in the perturbative limit for
a Wigner crystal in a magnetic field interacting with weak quenched disorder,
it has been argued
that the disorder will not affect the Hall angle \cite{Zhu94};
however, it is not clear what happens in the case of strong disorder.  

Depinning and sliding phenomena also appear in magnetic skyrmion
systems
\cite{Schulz12,Reichhardt15,Legrand17,Jiang17}, and 
it has been shown that the skyrmion dynamics
are very similar to those of electrons in a magnetic field
\cite{Schulz12,Nagaosa13}.
In particular,
the skyrmion motion exhibits a Hall angle 
with a value that depends on the materials parameters \cite{Nagaosa13}.
Simulations of skyrmion motion in the presence of disorder
reveal that the skyrmions have a finite depinning threshold
and that the skyrmion Hall angle is not constant but 
has a drive dependence, in which the Hall angle starts near
zero at depinning and increases with increasing drive
until reaching the disorder free limit at high drives
\cite{Reichhardt15,Legrand17,Nagaosa13,Kim17,Diaz17}.
This effect has now been observed directly in 
numerous imaging experiments
\cite{Jiang17,Litzius17,Woo18,Juge19,Zeissler20,Litzius20}.
Since skyrmions are extended bubble-like objects,
the current or disorder can distort
the shape or change the size of the skyrmions, potentially
generating the drive dependence of the skyrmion Hall angle
\cite{Litzius17,Juge19,Litzius20}.
Similar internal distortions 
could occur in a Wigner crystal.
Size distortions do not appear to be required, however, since
several
simulations using point-like approximations for the skyrmions
still found a strong drive dependence of the Hall angle
\cite{Reichhardt15,Diaz17,Gong20},
suggesting that this effect could be generic to other
particle-like systems exhibiting a Hall angle in the
presence of quenched disorder.
In the particle-based skyrmion model, the drive dependence
is the result of a velocity-dependent side jump effect,
where a particle that
passes through
a pinning site undergoes a jump
in the direction opposite to that of the Hall angle.
The jumps are more pronounced 
at lower velocity 
\cite{Reichhardt15,Muller15,Reichhardt15a}.
The presence of such jumps for skyrmions suggests
that a similar phenomenon could arise for any particle-like 
system with a Hall effect, such as a Wigner crystal in a magnetic
field,
and that the side jumps could be detected
by measuring changes in the Hall angle as a function
of drive. This would provide a new experimentally realizable method
for confirming the presence of a sliding Wigner crystal. 
    
We numerically examine a classical model for a Wigner crystal
interacting with random disorder using molecular dynamics simulations.
Previous studies of this model in the absence of a  magnetic field
showed that the system exhibits a depinning threshold and nonlinear
current-voltage (I-V) curves 
\cite{Cha94,Reichhardt01,Reichhardt04}.
When a magnetic field is applied, a driven Wigner crystal
in a pin-free system moves at a constant Hall angle.
If pinning is present, there
is a finite depinning threshold and the I-V curves become nonlinear.
In addition,
the Hall angle develops a strong drive dependence,
starting at a value near zero just above depinning
and saturating for high drives at a value
close to that found in the pin-free or
perturbative limit.
The drive dependence arises when a charge undergoes
a side jump effect upon interacting with the disorder sites. 
This effect can be confirmed by measurements
of the conduction both parallel and perpendicular to the drive.
It is predicted to be robust for a wide range of disorder strength and 
intrinsic Hall angles,
and it occurs for both a Wigner crystal and a Wigner glass.
The drive-dependent Hall effect
provides a new method to test for the
presence of a Wigner crystal in solid state systems.
Our results should also be general to 
other particle-like charged systems in quenched disorder 
and could be applied to charged colloids or dusty plasma systems
under a magnetic field.

{\it Simulation and System---} 
We consider the dynamics of a two-dimensional assembly of classical electrons
coupled to random pinning using molecular dynamics simulations. 
In the presence of a magnetic field, the
overdamped equation of motion for electron
$i$ is given by
\begin{equation} 
\alpha_d {\bf v}_{i} = \sum^{N}_{j}\nabla U(r_{ij}) + q{\bf B}\times {\bf v}_{i} +  {\bf F}_{p} + {\bf F}_{D} 
.
\end{equation}
Here $U(r)=q/r$ is the repulsive electron-electron interaction potential,
$q$ is the electron charge,
$q{\bf B}\times {\bf v}_{i}$ is the
force from the magnetic field which is oriented perpendicular to the
electron velocity ${\bf v}_i$, ${\bf F}_{p}$ is the
pinning force, ${\bf F}_{D}$ is the force from the externally applied drive,
and $\alpha_{d}$ is the damping term for the velocity component that is
aligned with the net external force direction.
The system contains $N_{e}$ electrons and $N_{\rm pin}$ pinning sites.
The electron density is given by $\rho=N_e/L^2$
and the pin density is $\rho_p=N_{\rm pin}/L^2$, where $L=36$ is the sample size.
The pinning is modeled as harmonic traps with a maximum force
of $F_{p}$ and radius $r_{p}$.
Here we consider the range $F_{p} = 0.0$ to $0.65$ with
$r_{p} = 0.35$.
There are periodic boundary conditions in the $x$ and $y$ directions,
and due to the long range nature of the electron-electron interactions,
we use a Lekner summation technique as in previous studies
\cite{Reichhardt01,Reichhardt04}.  

Previous work with this model was performed in the limit of no magnetic field,
$B=0$, where it was found 
that when pinning is present, there is a
finite depinning threshold for motion,
above which
the electrons move in the same direction
as the external driving force \cite{Reichhardt01,Reichhardt04}.
At finite $B$,
the combination of the damping
and driving causes the electrons to move at
a finite Hall angle $\theta_H$
with respect to the driving direction.
Since $\theta_H\propto\tan^{-1}(qB/\alpha_d)$, the Hall angle
increases with increasing $B$.
For convenience, we measure the magnetic field in units of
$B = \alpha_{d}/q$.
We apply a dc driving force ${\bf F}_{D}=F_D{\bf \hat{x}}$
and measure the net velocity both parallel and perpendicular
to the drive direction,
$\langle V_{||}\rangle = N^{-1}\sum^{N}_{i}{\bf v}_{i}\cdot {\hat {\bf x}}$ and
$\langle V_{\perp}\rangle = N^{-1}\sum^{N}_{i}{\bf v}_{i}\cdot {\hat {\bf y}}$,
giving a measured Hall angle of
$\theta_{\rm Hall} = \tan^{-1}(\langle V_{\perp}\rangle/\langle V_{||}\rangle)$.

\begin{figure}
\includegraphics[width=\columnwidth]{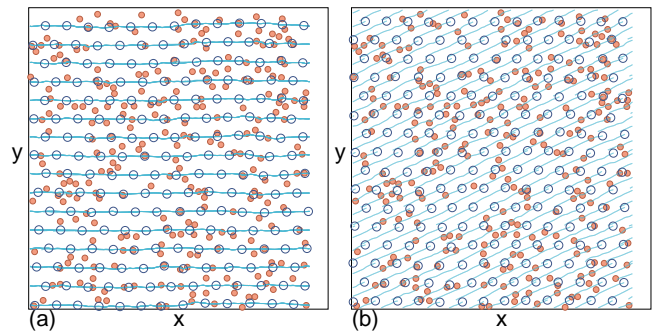}
\caption{Images of the Wigner crystal system showing
the electrons (blue open circles), pinning sites (orange filled circles),
and electron trajectories over a period of time (light blue lines).
The driving force $F_D=0.06$ is 
applied along the $x$-direction in both cases and
the pinning strength is $F_{p} = 0.1$.
(a) At zero magnetic field, $B=0$, the Hall angle $\theta_{\rm Hall}=0$.
(b) For a finite magnetic field of
$B=0.5$
where the intrinsic Hall angle is  $\theta_H=26.56^\circ$,
the Wigner crystal moves along a direction close to that of
the Hall angle. 
}
\label{fig:1}
\end{figure}

{\it Results-}
In Fig.~\ref{fig:1}(a) we show the electrons, pinning sites, and electron
trajectories for a system with
$F_{p} = 0.1$, $N_{e}/N_{\rm pin} = 0.803$,
electron density $\rho = 0.16$, and $B= 0.0$
at $F_{D} = 0.06$ where the drive is applied in the
$x$-direction.
A Wigner crystal forms and moves parallel to the driving
direction, giving
$\theta_{\rm Hall} = 0.0$.
Figure~\ref{fig:1}(b) shows the same system at a finite magnetic field where 
the intrinsic or disorder free Hall angle is  
$\theta_{H} = \tan^{-1}(q{B}/\alpha_d) =  26.56^\circ$.
Here the Wigner crystal moves at a finite angle
of $\theta_{\rm Hall} = 25^\circ$
which is slightly less than
the disorder free $\theta_H$.

\begin{figure}
\includegraphics[width=\columnwidth]{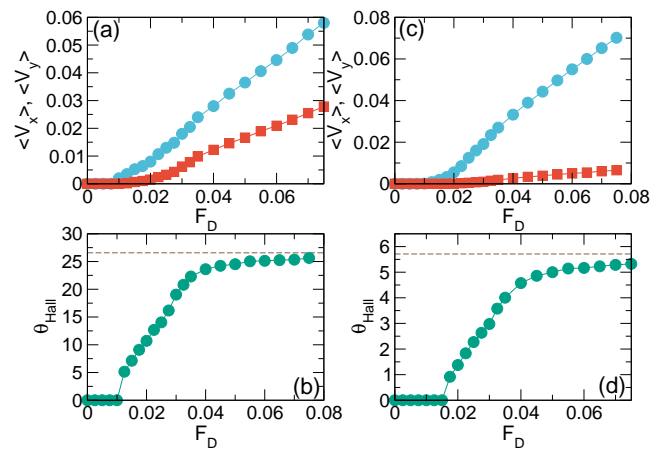}
\caption{ (a) The $x$ and $y$ velocities $\langle V_{x}\rangle$ (blue circles)
and $\langle V_{y}\rangle$ (red squares) vs $F_{D}$ for the
system in Fig.~\ref{fig:1}(b) with $F_p=0.1$ and
$B=0.5$.
(b) The corresponding Hall angle
$\theta_{\rm Hall} = \tan^{-1}(\langle V_{y}\rangle/\langle V_{x}\rangle)$.
The dashed line indicates
the disorder free value of $\theta_{H} = 26.56^\circ$.
(c) $\langle V_x\rangle$, $\langle V_y\rangle$ and (d) $\theta_{\rm Hall}$
vs $F_D$
for a system with a smaller magnetic field
of $B=0.1$ which gives
$\theta_{H} = 5.7^\circ$.
Here there is a low drive regime where the Hall angle increases
linearly with $F_D$ and a higher drive regime
where the increase in $\theta_{\rm Hall}$ with drive is much slower.
}
\label{fig:2}
\end{figure}

In Fig.~\ref{fig:2}(a) we plot
$\langle V_{x}\rangle$ and $\langle V_{y}\rangle$ versus $F_{D}$
for the system in Fig.~\ref{fig:1}(b), and
in Fig.~\ref{fig:2}(b) we show the corresponding 
$\theta_{\rm Hall}  = \tan^{-1}(\langle V_{y}\rangle/\langle V_{x}\rangle)$
versus
$F_{D}$.
A depinning transition appears near $F_{D} = 0.0075$,
and the nonlinear behavior of the velocity-force curve becomes
a linear behavior when
$F_{D} > 0.04$.
The dashed line in Fig.~\ref{fig:2}(b) indicates the pin-free
value of $\theta_{H}$, making it clear that
when quenched disorder is present, there is
an extended region of drive over which
$\theta_{\rm Hall}$ increases from zero and
then at higher drives begins to
saturate to a value close to the pin-free value.
In Fig.~\ref{fig:2}(c,d) we plot $\langle V_x\rangle$, $\langle V_y\rangle$,
and $\theta_{\rm Hall}$ versus $F_D$ for a sample with
a smaller magnetic field
giving $\theta_{H} = 5.7^\circ$,
where we observe the same behavior.
We find that for a wide range of
$\theta_{H}$, $F_{p}$, and $N_{\rm pin}$, 
the general trends illustrated in Fig.~\ref{fig:2} persist.
Namely, there is a strong drive dependence of the Hall angle
for drives up to 5 times the
depinning threshold, followed
by a crossover
to a saturation regime with a weaker drive dependence.

\begin{figure}
\includegraphics[width=\columnwidth]{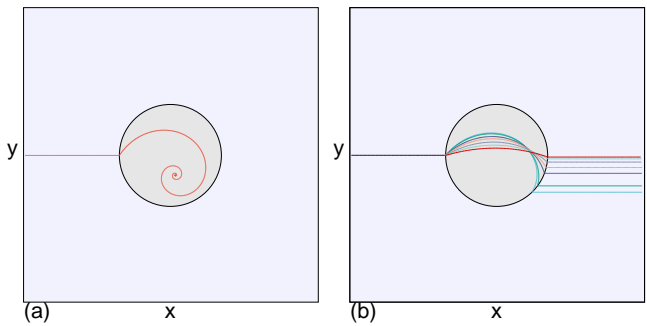}
\caption{ An electron
interacting with an attractive pinning well
that has $F_{p} = 0.5$.
Here the intrinsic Hall angle is
$\theta_H=75^\circ$ and the external drive is applied at an angle of
$-75^\circ$ to the $x$ axis such that, in the pin-free limit,
the electron moves parallel to the $x$ axis.
(a) At $F_{D} = 0.2$, the electron becomes trapped in the pin.
(b) $F_{D} = 0.275$ (light blue), 0.3 (dark green), 0.4 (dark blue),
0.5 (orange), 0.7 (light purple), 1.0 (light green), and 1.5 (red) showing
the evolution of the side jump as the electron passes across the pinning site
for drives above the depinning threshold. The side jump is in the direction
opposite to the Hall angle, and the size of the side jump
decreases as the driving force increases.
}
\label{fig:3}
\end{figure}

The velocity dependence of the Hall angle is the result of a side jump effect 
that occurs as the charges move though the pinning sites. 
To illustrate this effect, in Fig.~\ref{fig:3} we
plot the trajectories of a single electron interacting with a pinning site
with $F_p=0.5$.
To better highlight the
side jumps 
we use a large intrinsic Hall angle of
$\theta_H=75^{\circ}$, and we apply the external drive at an
angle of $-75^{\circ}$ from the $x$ axis so that in the absence of pinning
the electron motion would be parallel to the $x$ axis.
For $F_{D} = 0.2$ the electron is trapped by the
pinning site and undergoes a spiraling motion to
reach the minimum energy position of the tilted pin, as shown in 
Fig.~\ref{fig:3}(a).
For drives that are large enough to permit the electron to escape the trap,
the electron trajectory is bowed across the pinning site and the point of
exit of the electron is shifted in the negative $y$ direction compared to
the point of entry.
Figure~\ref{fig:3}(b) illustrates this effect for 
$F_{D} = 0.275$, 0.3, 0.4, 0.5, 0.7, 1.0, and 1.5,
where the shift or jump gradually decreases in size with increasing
$F_{D}$.
The  side jumps are in the direction opposite to the Hall angle, so that
for repeated interactions with pinning sites,
the overall average electron motion is at
an angle that is smaller than the intrinsic Hall angle $\theta_H$.
As the velocity of the
electron increases, its direction of motion approaches
$\theta_H$. 

\begin{figure}
\includegraphics[width=\columnwidth]{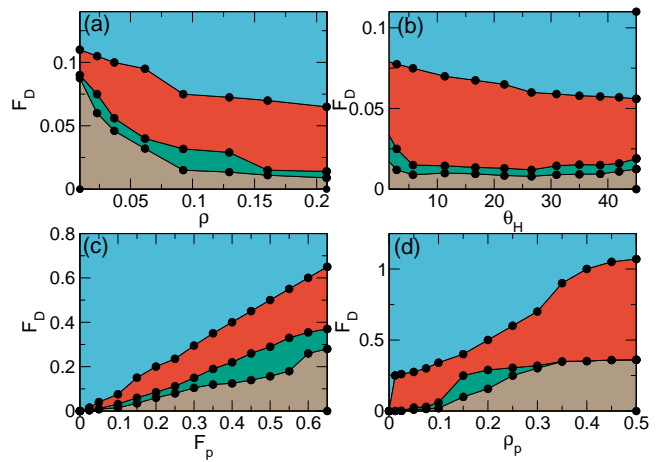}
\caption{ (a) Dynamic phase diagram as a function of $F_D$ vs
electron density $\rho$ for the system in Fig.~\ref{fig:2} with
$F_p=0.1$, pin density $\rho_p=0.2$, 
and $B=0.1$, where the intrinsic Hall angle is $\theta_H=5.7^\circ$.
Colors indicate the pinned regime (brown), 
the flowing saturated regime (blue),
the linearly increasing Hall angle regime (orange)
and the sliding but zero Hall angle state (green). 
(b) Dynamic phase diagram as a function of $F_D$ vs intrinsic
Hall angle $\theta_H$ for the same system at fixed $\rho=0.16$.
(c) Dynamic phase diagram as a function of $F_D$ vs pinning strength
$F_p$ for a system with $\rho = 0.0926$, $\rho_p=0.2$,
and $\theta_{H} = 5.7^\circ$.
(d) Dynamic phase diagram as a function of $F_{D}$ vs
pinning density $\rho_{p}$ for a sample with $\theta_{H} = 5.7^\circ$,
$\rho = 0.0926$, 
and $F_{p} = 0.5$.  
}
\label{fig:4}
\end{figure}

In Fig.~\ref{fig:4}(a) we plot a dynamic phase diagram as a function of
$F_D$ versus electron density $\rho$
for the system in Fig.~\ref{fig:2}(c,d)
at fixed pinning density $\rho_p=0.2$ for
$\theta_{H} = 5.7^\circ$.
We highlight the
pinned state, the flowing state in which the Wigner crystal moves nearly
along the intrinsic Hall angle direction,
the regime in which the Hall angle increases linearly with drive,
and the region where the crystal is moving but the Hall angle is zero.
The latter regime is likely the result of
a transverse barrier to motion.
Such a transverse pinning effect can also arise in the
absence of a magnetic field,
as was proposed by Giamarchi and Le Doussal \cite{Giamarchi96,LeDoussal98}
for vortices and other particle like
objects moving over random disorder where the system forms
one-dimensional moving channels.
Previous numerical studies
on sliding Wigner crystals in the absence of a magnetic field also
found a transverse barrier for motion when an additional
driving force was applied transverse to the sliding
direction \cite{Reichhardt01}.
In the present case, there is no additional applied transverse drive,
but there is a finite Hall angle.
In Fig.~\ref{fig:4}(a), for $\rho < 0.075$, the system forms a pinned Wigner
glass rather than a pinned Wigner crystal, resulting in an increase 
in the depinning threshold with decreasing $\rho$.
Figure~\ref{fig:4}(b) shows a dynamic phase diagram as a function of
$F_D$ versus the intrinsic Hall angle $\theta_H$ over the range
$\theta_H=2.86^\circ$ to $45^\circ$
for the same system at fixed $\rho=0.16$.
In Fig.~\ref{fig:4}(c) we display the dynamic phase diagram 
as a function of $F_{D}$ versus pinning strength $F_{p}$
for a system with $\theta_{H} = 5.7^\circ$ and $\rho = 0.0926$,
while in Fig.~\ref{fig:4}(d) we plot a dynamic phase diagram as a function of
$F_D$ versus pinning density $\rho_p$
for a system with fixed $\rho = 0.0926$, $F_{p} = 0.5$,
and $\theta_{H}= 5.7^\circ$. These results indicate that the dynamic phases
we observe
are robust over a wide range of parameters.

Experimentally, the drive dependence of the Hall angle
could be detected by measuring the transport above a depinning threshold
in a system where a Wigner crystal is expected
to form in the presence of a magnetic field. 
The drive dependence should be the most pronounced for
large magnetic fields.
There have already been
some limited experimental studies 
of Wigner crystal sliding in a magnetic field which
show that there is a minimum threshold longitudinal velocity
which must be exceeded
before sliding begins to occur
in the transverse direction \cite{Perruchot00}.
Other systems in which
a similar effect could appear
include a sliding quantum crystal \cite{Chen06,Shashkin19},
creep motion of a Wigner crystal 
near melting \cite{Knighton18,Ma20} or
driven electron liquid crystal states
\cite{Cooper99,Reichhardt03a,Wang15}.
In some regimes in these systems,
the particle picture breaks down,
and in these regimes,
the drive dependence of the Hall effect could be absent.
Additionally,
a similar drive dependent Hall effect
should occur for charged colloids or dusty plasmas
driven in the presence of quenched disorder and a magnetic field. 

{\it Summary---} 
We have examined the depinning and sliding of a Wigner crystal
in the presence of a magnetic field where in the absence of 
disorder the crystal moves at a Hall angle that is independent of the
crystal velocity.
When disorder is present, we find a pinned phase at low drive as well as a
sliding phase in which the Hall angle is not constant
but is initially zero near the depinning threshold and gradually increases
with increasing drive until at high drives it reaches a saturation value
close to the intrinsic Hall angle.
This effect is the result of
the side jump that occurs
when the electrons move over the pinning sites,
where the jump is in the direction opposite to the Hall angle.
The magnitude of the side jump decreases with increasing
velocity, which is similar to the behavior
of the side jump effect observed for skyrmions driven over quenched disorder. 
We also find a regime in which the
electrons slide only along the direction of drive and
exhibit a finite barrier to transverse motion.
The drive dependence of the Hall angle is robust
over a wide range of intrinsic Hall angles, disorder densities, and
pinning strengths, and could serve as a new way to confirm the existence
of a sliding Wigner crystal. 
Our results should be general to the broad class of driven systems
with a Hall angle in the presence of disorder. 

\begin{acknowledgments}
We gratefully acknowledge the support of the U.S. Department of
Energy through the LANL/LDRD program for this work.
This work was supported by the US Department of Energy through
the Los Alamos National Laboratory.  Los Alamos National Laboratory is
operated by Triad National Security, LLC, for the National Nuclear Security
Administration of the U. S. Department of Energy (Contract No. 892333218NCA000001).
\end{acknowledgments}

\bibliography{mybib}

\end{document}